\documentclass[twocolumn,showpacs,amsmath,amssymb,prl]{revtex4}

\usepackage{graphicx}

\begin{document}

\title{Neutrinos from Pulsar Wind Bubbles as Precursors to Gamma-Ray Bursts}

\author{Jonathan Granot$^\dagger$ and Dafne Guetta$^\ddagger$}

\affiliation{$^\dagger$Institute for Advanced Study, Princeton, NJ
08540; E-mail: granot@ias.edu\\
$^\ddagger$Osservatorio astrofisico di Arcetri, 
Fermi 5, 50125 Firenze, Italy; E-mail: dafne@arcetri.astro.it}

\date{\today}

\begin{abstract}
The supranova model for $\gamma$-ray bursts (GRBs) has recently gained popularity.
In this scenario the GRB occurs weeks to years after a supernova explosion, and is 
located inside a pulsar wind bubble (PWB). High energy protons from the PWB can 
interact with photons from the rich radiation field inside the PWB  
or collide with cold protons from the supernova remnant, producing pions which decay 
into $\sim 10-10^3\,\,$TeV neutrinos. The predicted neutrino flux from the PWBs that 
host the GRBs should be easily detectable by planned ${\rm 1\; km^2}$ detectors.
\end{abstract}

\pacs{96.40.Tv, 98.70.Rz, 98.70.Sa}
\maketitle


The leading models for Gamma-Ray Bursts (GRBs) involve a relativistic 
outflow emanating  
from a compact source. The $\gamma$-rays are 
attributed to synchrotron (and possibly also inverse-Compton)
emission from relativistic electrons 
accelerated in internal shocks within the ejecta.
The ultimate energy source is 
rapid accretion onto a newly formed stellar mass black hole.
Long duration ($\gtrsim 2\;{\rm s}$) GRBs, which include all GRBs with observed 
afterglow emission and $\sim 2/3$ of all GRBs, are widely assumed to originate 
from a massive star progenitor. This is supported by growing evidence for GRBs 
occurring in star forming regions within their host galaxies~\cite{SFR}.
The leading model for long duration GRBs is the collapsar model~\cite{coll},
where a massive star promptly collapses into a black hole, and forms a 
relativistic jet that penetrates through the stellar envelope and produces 
the GRB. An interesting alternative, though somewhat more debated, model is the
supranova model~\cite{supra}, where a supernova explosion leaves behind a 
supra-massive neutron star (SMNS), of mass $\sim 2.5-3\,M_\odot$, which loses 
its rotational energy on a time scale, $t_{\rm sd}\sim\;$weeks to years, and 
collapses to a black hole, triggering the GRB. The most natural mechanism by 
which the SMNS can lose its rotational energy is through a strong pulsar type 
wind, which is expected to create a pulsar wind bubble (PWB)~\cite{KGGG,GG}.

In this Letter we consider the neutrino production in the PWB 
in the context of the supranova model for GRBs. 
These $\nu$'s have $\sim 10\;$TeV energies and 
are emitted over the time $t_{\rm sd}$ between the supernova and GRB events, 
for $t_{\rm sd}\lesssim 1\;$yr, while for $t_{\rm sd}\gtrsim 20\;$yr, 
that is required for GRBs with typical afterglows 
(for spherical PWBs)~\cite{GG}, they are emitted mainly over the first few 
years after the supernova. For $t_{\rm sd}\lesssim 0.1\;$yr
or for GRBs not pointed towards us, 
the $\nu$'s would not be
accompanied by a detectable GRB.
We find that the $\sim 10\;$TeV neutrino 
fluence from a PWB at $z\sim 0.1-1$ implies $\sim 0.1-10$ upward moving muons 
in a ${\rm km}^{2}$ detector. We expect $\sim 10$ nearby ($z\lesssim 0.1$)
PWBs per year. The neutrino signal from an individual PWB for which the GRB 
jet is pointed towards us, and is therefore detectable in $\gamma$-rays, 
is above the atmospheric neutrino background.  

\paragraph*{Neutrino production in the PWB.}---
The main mechanisms for neutrino production are (1)
photomeson interactions between relativistic protons 
and nonthermal photons inside the PWB ($p\gamma$), (2) p-p collisions 
between the PWB protons and cold SNR protons ($pp$), (3) p-p collisions 
between shock accelerated SNR protons and cold SNR protons ($sh$); 
p-p collisions between two relativistic protons inside the PWB are unimportant. 
The pulsar wind is expected to have a comparable energy in protons and in
$e^\pm$ pairs. The proton velocities are randomized at the wind termination 
shock, and are expected to be quasi-monoenergetic, with a random 
Lorentz factor $\gamma_p=10^{4.5}\gamma_{p,4.5}$ (i.e. an energy 
$\epsilon_p=30\gamma_{p,4.5}\;$TeV) since this is the value needed to explain 
GRB afterglow observations with $t_{\rm sd}\gtrsim 20\;$yr 
(for a spherical PWB)~\cite{GG}. 
However, $\gamma_p$ can be larger for 
$t_{\rm sd}=0.1t_{\rm sd,-1}\;{\rm yr}<20\;$yr. 

The fraction, $\chi_p$, of PWB protons that can reach the SNR  
and may interact with SNR protons, is uncertain. The ratio of the 
proton Larmor radius, $R_{L}$, and the SNR radius, $R_b$, is 
$\sim 4\cdot 10^{-7}\gamma_{p,4.5}\xi_{B,-3}^{-1/2}t_{\rm sd,-1}^{1/2}\ll 1$,
where $\xi_{B}=10^{-3}\xi_{B,-3}$ is the fraction of PWB energy in magnetic 
fields, and the protons should largely follow the magnetic field lines. 
Polarization maps of the Crab nebula~\cite{crab} suggest that the 
magnetic field in plerions is ordered on scales $\sim R_b$.
This may suggest a reasonably high $\chi_p$. Plasma instabilities can 
reduce $\chi_p$, while a radial flow along the poles (jets) can 
increase $\chi_p$. Therefore, the value of $\chi_p$ can range anywhere
from $\sim 1/2$ to $\ll 1$.

The average optical depth for p-p collisions between a proton from the PWB, 
that propagates in the radial direction, and the protons 
of the SNR shell is
\begin{equation}\label{tau_pp}
\tau_{pp}^{\rm rad}={\sigma_{pp}M_{\rm SNR}\over 4\pi R_b^2 m_p}
=1.2\left({M_{\rm SNR}\over 10\,M_\odot}\right)
\beta_{b,-1}^{-2}t_{\rm sd,-1}^{-2}\ ,
\end{equation}
where $M_{\rm SNR}$ and $\beta_b=0.1\beta_{b,-1}$ are the mass and the 
velocity (in units of $c$) of the SNR shell, and 
$\sigma_{pp}\approx 50\,{\rm mb}$. The typical optical depth 
to photomeson interactions is 
$\tau_{p\gamma}\approx 0.14\gamma_{p,4.5}^{1.3}\tau_{pp}^{\rm rad}$,
so that $\tau_{p\gamma}\gtrsim\tau_{pp}^{\rm rad}$ for $\gamma_{p}\gtrsim 10^5$.
The rotational energy lost by the SMNS prior to its 
collapse, $E_{\rm rot}=10^{53}E_{53}\;$erg, is comparable to its total 
rotational energy~\cite{supra}.
Typically $M_{\rm SNR}\sim 10\,M_\odot$, implying $\beta_{b}\sim 0.1$~\cite{supra}. 
Even though initially $\beta_b$ is probably $\ll 0.1$, 
the large pressure inside the PWB accelerates the SNR to a velocity such 
that its kinetic energy becomes $\sim E_{\rm rot}$~\cite{KGGG,GG}.
During the acceleration, Rayleigh-Taylor instabilities are expected to 
develop~\cite{GG,IGP} that may condense the SNR shell into clumps or filaments, 
as is seen in the Crab nebula. This increases the interface between the PWB 
and the now fragmented SNR, which helps increase $\chi_p$.

The interaction with the pulsar wind drives a collisionless
shock into the SNR shell~\cite{RMW}, that crosses it on a time 
$\sim t_{\rm sd}$, and can accelerate protons up to energies 
$\epsilon_{p,{\rm max}}\sim 10^{16}\;$eV,
with a roughly flat $\epsilon_p^{2}(dN/d\epsilon_p)$ energy spectrum.
These protons can collide with cold SNR protons, providing an additional channel 
for neutrino production. The neutrino energy spectrum will follow that 
of the protons, where  
$\epsilon_\nu\approx\epsilon_p/20$.
The total energy in these protons is a fraction $f_p\sim 0.1-0.5$ 
of $E_{\rm rot}$.

The pion luminosities 
in the different channels are
\begin{equation}\label{L_pi_pp}
{L_{\pi}^{p\gamma}\over \xi_p f_{\pi}^{p\gamma}}=
{L_{\pi}^{pp}\over\chi_p \xi_p f_{\pi}^{pp}}=
{L_{\pi}^{sh}\over f_p f_{\pi}^{pp}}=
3.2\times 10^{46}{E_{53}\over
t_{\rm sd,-1}}\,{\rm {erg\over s}}\ ,
\end{equation}
where $\xi_p=(2/3)\xi_{p,2/3}$ is the 
fraction of wind energy in protons,  
and $f_{\pi}^{pp}$ 
($f_{\pi}^{p\gamma}$) is the fraction of the total proton energy 
lost to pion production through p-p collisions 
(p-$\gamma$ interactions). 
A fraction $\eta_p\approx 0.2$ of the proton energy is lost to pion 
production in a single collision (interaction). 
The energy loss in multiple collisions can be approximated by a continuous 
process, where $\epsilon_p=\epsilon_{p,0}\exp(-\tau^{\epsilon}_{i})$.
On average, $n=-\ln(1-\eta_p)^{-1}\approx 4.5$ collisions produce one e-folding, 
and $\tau^{\epsilon}_{i}=\tau_{i}/n\approx\tau_{i}/4.5$, where 
$i=pp,\,p\gamma$. This implies 
$f_{\pi}^{i}\approx 1-\exp(-\tau^{\epsilon}_{i})\approx 1-\exp(-0.22\tau_{i})$.

The shock going into the SNR shell can produce a tangled magnetic field 
with a strength similar to that inside the PWB.
Thus, once a PWB proton enters the SNR shell, it is likely to stay there
over a time $\gtrsim t_{\rm sd}$. This can increase the effective 
value of $\tau_{pp}$ by up to a factor of $\sim[\beta_b(\Delta R/R_b)]^{-1}\sim 100$,
compared to $\tau_{pp}^{\rm rad}$ (Eq. \ref{tau_pp}).
Therefore,  $\tau_{pp}\gtrsim 1$ and $f_\pi^{pp}\sim 1$ 
for $t_{\rm sd}\lesssim 1\;$yr, $\tau_{p\gamma}\gtrsim 1$ 
and $f_\pi^{p\gamma}\sim 1$ 
for $t_{\rm sd}\lesssim 0.04\gamma_{4.5}^{0.65}\;$yr, and 
$\tau_{pp}\propto t_{\rm sd}^{-2}$, $\tau_{p\gamma}\propto t_{\rm sd}^{-2.1}$. 
We have $\tau_{p\gamma}\sim\tau_{pp}$ for $\gamma_p\sim 10^6-10^7$.

When $\tau_{p\gamma}>1$, the PWB protons lose all their energy 
via photomeson interactions before they can reach $R_b$
and collide with the SNR protons. When $\tau_{p\gamma}\sim 1$,
$L_\pi^{pp}\sim L_\pi^{p\gamma}$, since the protons lose only $\sim 20\%$ 
of their energy in a single $p\gamma$ interaction, and in half of the 
cases are converted to neutrons, that can reach the SNR (since they are not 
effected by magnetic fields, and as long as 
$\gamma_p>R_b/c\tau_n\sim\beta_bt_{\rm sd}/\tau_n$, where $\tau_n\approx 900\,$s 
is the neutron mean lifetime) carrying $\sim 1/2$ of the PWB proton energy, 
and collide with the protons there 
(typically $\tau_{pp}\gtrsim\tau_{p\gamma}$). 

So far, all the parameters were estimated at the time of the GRB, 
$t=t_{\rm sd}$. However, since $\tau_{i}\propto t^{-a}$ where $a_{pp}=3.5$ 
($\tau_{pp}\propto t/R^3$) and $a_{p\gamma}=2.2$~\cite{GG}, 
even if $\tau_{i}(t_{\rm sd})<1$, at early times,
$t\lesssim t_{i}=t_{\rm sd}\tau_{i}(t_{\rm sd})^{1/a}$,
$\tau_{i}>1$ and $f_\pi^{i}\sim 1$. The mean value of $f_\pi^{i}$ over 
$0<t\lesssim t_{\rm sd}$ is $\sim\tau_{i}(t_{\rm sd})^{1/a}a/(a-1)$, 
and most of the neutrinos are emitted at $0<t\lesssim t_{i}$.

The pions decay on a very short time scale, before they
can suffer significant energy losses.
The pion energy 
is divided roughly
$1:1:1$ ($1:2:1$) between $\nu_\mu\bar\nu_\mu$, $\gamma$-rays
and other products, for p-p (p-$\gamma$) reactions. 
The initial flavor ratio 
$\Phi_{\nu_e}:\Phi_{\nu_\mu}:\Phi_{\nu_\tau}$ is $1:2:0$, however, 
due to neutrino oscillations we expect a ratio of  
$1:1:1$ at the Earth. 
The $\nu_\mu\bar\nu_\mu$ fluence over the lifetime of the PWB is
\begin{equation}\label{fluence}
{8f_{\nu_{\mu}}^{p\gamma}\over\xi_p f_{\pi}^{p\gamma}}=
{6f_{\nu_{\mu}}^{pp}\over\chi_p \xi_p f_{\pi}^{pp}}=
{6f_{\nu_{\mu}}^{sh}\over f_p f_{\pi}^{pp}}=
{8.0\times 10^{-5}E_{53}\over d_{L28}^{2}/(1+z)}\,{\rm erg\over cm^{2}}\ ,
\end{equation}
where $z$ and $d_L=10^{28}d_{L28}\;$cm are the cosmological redshift and
luminosity distance of the PWB, respectively. The average flux is just the 
fluence divided by $t_{\rm sd}$, and the fluence in $\gamma$-rays is  
$f_\gamma^{p\gamma}=4f_{\nu_\mu}^{p\gamma}$, $f_\gamma^{pp}=2f_{\nu_\mu}^{pp}$,
$f_\gamma^{sh}=2f_{\nu_\mu}^{sh}$.
 
The average Thomson optical depth of the SNR shell is 
$\langle\tau_T\rangle=(\sigma_T/\sigma_{pp})\tau_{pp}^{\rm rad}
\approx 13\tau_{pp}^{\rm rad}\approx(t_{\rm sd}/0.4\,{\rm yr})^{-2}$. 
Thereby $f_{\pi}^{pp}\sim 1$ requires
$\tau_{pp}\sim 100\tau_{pp}^{\rm rad}\gtrsim 1$ and 
$\langle\tau_T\rangle\gtrsim 0.1$, while $f_{\pi}^{p\gamma}\sim 1$ 
requires $\langle\tau_T\rangle\gtrsim 100\gamma_{p,4.5}^{-1.3}$.
If $\tau_T>1$, low energy photons cannot escape. 
High energy photons, above the Klein-Nishina limit ($\epsilon_\gamma>m_e c^2$)
have a reduced cross section and can escape if 
$\epsilon_\gamma\gtrsim \tau_T m_e c^2\approx 5t_{\rm sd,-1}^{-2}\;$MeV.
Since $\epsilon_\gamma$ is typically $\lesssim m_e c^2$ in the prompt GRB, 
it would not be detectable for a uniform SNR with $t_{\rm sd}\lesssim 0.4\;$yr.
The synchrotron self-Compton GRB emission
could still escape. 
However, the optical depth of high energy GRB photons to pair 
production with the low energy PWB photons is $\tau_{\gamma\gamma}\gtrsim 1$
for $\epsilon_\gamma\gtrsim 30t_{\rm sd,-1}^2\;$MeV~\cite{GG}.
This leaves a narrow range of photon energies that can 
escape the PWB, $\epsilon_\gamma\sim 5-30\;{\rm MeV}$,
for the parameter range where neutrino 
production is most efficient ($t_{\rm sd,-1}\lesssim 1$).
If the SNR shell is clumpy, then $\tau_T\ll\langle\tau_T\rangle$ 
at the under-dense regions, relaxing the constraints related to $\tau_T$.
Moreover, $f_\pi^{pp}\sim 1$ for $t_{\rm sd}\lesssim 1\;$yr, 
so that efficient neutrino production 
via p-p collisions can occur while all photons with 
$\epsilon_\gamma\lesssim 3(t_{\rm sd}/1\,{\rm yr})^2\;$GeV
can escape for $1\lesssim t_{\rm sd,-1}\lesssim 10$ 
(or $4\lesssim t_{\rm sd,-1}\lesssim 10$ for a uniform shell).

Detecting the GRB emission helps distinguish between
$\nu'$s from the PWB and $\nu'$s from the atmospheric background, since the 
PWB $\nu'$s arrive from the same direction and are correlated in time
with the GRB photons.

The $\pi^0$'s that are produced 
decay into 
$\epsilon_{\gamma,0}\sim\epsilon_p/10\sim 3\gamma_{p,4.5}\;$TeV photons. 
These high energy photons will produce pairs with the low energy PWB 
photons. The high energy $e^\pm$ pairs will, in turn, upscatter 
low energy photons to high energies, etc., thus creating a pair cascade. 
This enables some photons to 
escape the system, even if initially 
$\tau_{\gamma\gamma}(\epsilon_{\gamma,0})>1$, after 
having overgone $\sim\log_{2}[\tau_{\gamma\gamma}(\epsilon_{\gamma,0})]$ 
scatterings, and shifted down to an energy 
$\epsilon_\gamma\sim\epsilon_{\gamma,0}/\tau_{\gamma\gamma}(\epsilon_{\gamma,0})$, 
such that 
$\tau_{\gamma\gamma}(R_b)\lesssim 1$ and 
$\tau_T\min(1,m_e c^2/\epsilon_\gamma)\lesssim 2$. 
Since $\tau_{\gamma\gamma}(R_b)\lesssim 1$ for 
$\epsilon_\gamma\gtrsim 90t_{\rm sd,-1}^2\;$MeV, then for 
$2\lesssim\tau_T\lesssim 180t_{\rm sd,-1}^2$
the escape of the photons is limited by $\tau_{\gamma\gamma}(R_b)$, 
and the photons that escape have a typical energy
$\epsilon_\gamma\sim 90t_{\rm sd,-1}^2\;$MeV. 
These photons should accompany
the high energy neutrinos.

When the SMNS collapses to a black hole 
and triggers the GRB, the pulsar wind stops abruptly.
However, the relativistic protons in the PWB or SNR can still produce pions,
and the neutrino emission via p-p collisions would not stop abruptly 
after the GRB, but would rather decay over the expansion time, 
$\sim t_{\rm sd}$, over which the hot protons lose most of their energy 
via adiabatic cooling, and $\tau_{pp}$ decreases. 
Neutrino emission via photomeson interactions typically decays on
a shorter time scale since the PWB radiation field is produced by 
fast cooling electrons.

The neutrino spectrum (for each of the 3 $\nu$ flavors) for a mono-energetic proton 
distribution is given by~\cite{Blasi}
\begin{eqnarray}\label{dN_dE_nu_pp}
\Phi_\nu &\equiv&{dN_\nu\over d\epsilon_\nu dAdt}=
K\, {1\over 4\epsilon_\nu}\, 
g_\pi\left({4\epsilon_\nu\over \epsilon_p}\right)\ ,
\\ \label{A_nu}
K &=& \left\{\begin{matrix} (1/2)L_\pi^{p\gamma} \vspace{0.12cm} \cr
      (2/3)L_\pi^{pp} \end{matrix}\right\}\times
{(1+z)^2 \over \pi\epsilon_p d_L^2}\left(\int_0^1g_\pi(x)dx\right)^{-1}
\\ \nonumber
&=& \left\{\begin{matrix} 2.7 f_\pi^{p\gamma} \vspace{0.12cm} \cr
3.6 \chi_p f_\pi^{pp} \end{matrix}\right\}\times 
{\xi_{p,2/3}E_{53}(1+z)^2
\over\gamma_{p,4.5}t_{\rm sd,-1}d_{L28}^{2}}
\;{10^{-12}\over{\rm s\;cm^{2}}}\ ,
\end{eqnarray}
where $g_\pi(x)=(1-x)^{3.5}+\exp(-18x)/1.34$, and $\epsilon_\nu$ ranges between 
$m_\pi c^2/4\approx 25\;$MeV (or $\sim\gamma_{\rm cm}m_\pi c^2/4
\sim 3\gamma_{p,4.5}^{1/2}\;$GeV for p-p collisions)
and $\epsilon_p/4\approx 7.4\gamma_{p,4.5}\;$TeV.
The normalizations for $p\gamma$ and for $pp$ are given separately.
The original spectrum of the $\gamma$-rays produced in the $\pi^0$ decay should 
be roughly similar to that of the neutrinos.

The probability of detecting a muon neutrino in a terrestrial
ice detector is 
$P_{\nu\mu}\approx 1.3\times 10^{-6}(\epsilon_\nu/{\rm TeV})^\beta$, with 
$\beta=2$ for $\epsilon_\nu<1\;$TeV and $\beta=1$ for 
$\epsilon_\nu>1\;$TeV~\cite{nu_det}.
For $\epsilon_\nu>10^{15}\;$eV, $P_{\nu\mu}\propto\epsilon_\nu^{1/2}$.
The expected number of neutrino events during the life time of the PWB is 
\begin{eqnarray}\label{N_nu_PWB}
N_{\mu}&=&  A\,\min\left[(1+z)t_{\rm sd},T\right] 
\int\Phi_{\nu_\mu}(\epsilon_\nu)P_{\nu\mu}(\epsilon_\nu)d\epsilon_\nu\ ,\ \ \ 
\\
N_{\mu}^{p\gamma} &=& 0.05\, f_{\pi}^{p\gamma}\xi_{p,2/3}E_{53}
(1+z)d_{L28}^{-2}(A/{\rm km^2})\ ,
\\ 
N_{\mu}^{pp} &=& 0.07\,f_{\pi}^{pp}\chi_p\xi_{p,2/3}E_{53}
(1+z)d_{L28}^{-2}(A/{\rm km^2})\ ,
\\
N_{\mu}^{sh} &=& 0.05\,f_{\pi}^{pp} f_p E_{53}(1+z)d_{L28}^{-2}(A/{\rm km^2})\ ,
\end{eqnarray}
where $A$ and $T\sim 1\;$yr are the area and integration time of 
the detector, and we have assumed $T>(1+z)t_{\rm sd}$.
These expected numbers of events are valid when absorption in the Earth 
is not important, and apply to GRBs that point at
the detector horizontally, and to $\epsilon_\nu<100\;$TeV neutrinos
from GRBs at the hemisphere opposite of the detector.
Absorption may reduce the number of $\epsilon_\nu>100\;$TeV events,
which are expected only for $\gamma_p\gtrsim 10^6$.

\paragraph*{Implications.}

The atmospheric neutrino background flux is $\Phi_\nu^{\rm bkg}\sim 10^{-7}
(\epsilon_\nu/{\rm TeV})^{-2.5}\;{\rm cm^{-2}\,s^{-1}\,sr^{-1}}$.
The angular resolution, $\theta$, of the planned neutrino telescopes ICECUBE 
and NEMO will be $0.7^\circ$~\cite{ICECUBE} and $0.3^\circ$~\cite{NEMO}, 
respectively. The number of background
events above a neutrino energy $\epsilon_\nu$, in the range $\sim 1-10^3\;$TeV,
is $N_{\rm bkg}(>\epsilon_\nu)\sim 10^{-2}(\theta/0.3^\circ)^{2}
(\epsilon_\nu/10\,{\rm TeV})^{-1.5}(t/0.1\,{\rm yr})$ per angular resolution 
element, over a time $t$. There is better signal to noise ratio for larger 
$\epsilon_p=\gamma_p m_pc^2$, which are possible for the parameter range 
where neutrino emission is most efficient ($t_{\rm sd}\lesssim 1\;$yr).
Thus, the neutrino signal from a single PWB with 
$t_{\rm sd}\sim 0.1\;$yr, at $z\sim 1$, could be detected with high 
significance above the background. 

GRBs that occur inside spherical PWBs with
$t_{\rm sd}\sim 0.1-1\;$yr would have a peculiar and short lived afterglow 
emission~\cite{GG} (such events may be related to X-ray flashes, for which no
afterglow emission was detected so far).
Since they occur for a wide range of 
$t_{\rm sd}$ values, we expect their rate to be similar to that of typical
GRBs (i.e. $\sim 10^3\;{\rm yr^{-1}}$ that are beamed towards us and 
$\sim 10^5\;{\rm yr^{-1}}$ that are beamed away from us).
Therefore, a ${\rm km^2}$ neutrino detector should detect $\sim 100\;{\rm yr^{-1}}$ 
neutrinos correlated with GRBs. 
Furthermore, $\sim 10\;{\rm yr^{-1}}$ close by ($z\lesssim 0.1$) PWBs
are expected, where each PWB will produce at least several events, 
and would thereby be above the atmospheric background, even if 
the GRB cannot be detected in $\gamma$-rays, as it is beamed away 
from us or since $\tau_T>1$ for $t_{\rm sd}\lesssim 0.1\;$yr.
In the limit of very small 
$t_{\rm sd}$ ($\lesssim \;$a few hours), this reduces to the 
chocked GRBs in the collapsar model~\cite{coll_nu}.
 
\begin{figure}[!t]
\includegraphics[width=3.3in]{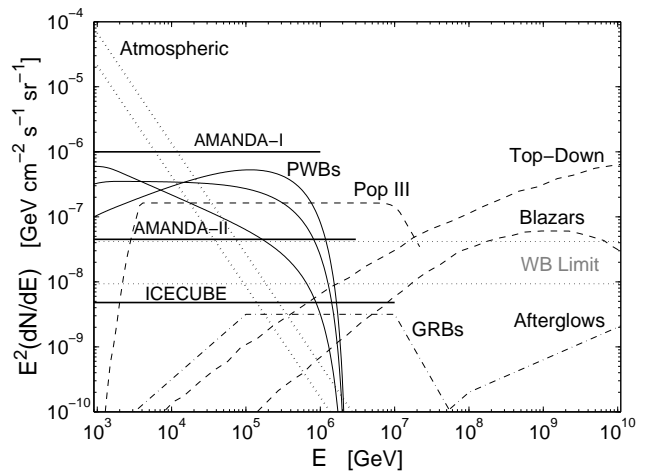}
\caption{\label{fig1}
The diffuse neutrino flux from different sources is compared to 
the sensitivities of present and future telescopes.
The expected flux from PWBs that host GRBs, due to the $pp$ and $p\gamma$
mechanisms (see text for details), is shown for 3 
different distributions of $\gamma_p$ among the different PWBs 
({\it thin solid lines}): 
$d{\rm Pr}/d\log(\gamma_p)\propto\gamma_p^\alpha$, $10^4\leq\gamma_p\leq 10^7$, 
$\alpha=-0.5,\,0,\,0.5$. The dotted lines show the Waxman-Bahcall cosmic-ray limit.
}
\end{figure}

The diffuse muon neutrino flux from PWBs that host GRBs
is shown in Fig.~1, for the $p\gamma+pp$ mechanisms, 
using the flux normalization from Eq. (\ref{A_nu}) with
$f_\pi^{p\gamma}=f_\pi^{pp}=2\chi_p=1$,
and a rate of $10^5\;{\rm yr^{-1}}$ at $z\sim 1$.
It is slightly below the existing upper bound on the diffuse flux 
established by AMANDA~\cite{AMANDA}, and may be detected by AMANDA II.
It is above the Waxman-Bahcall (WB) cosmic ray limit~\cite{WBlimit}.
This is not a problem as these are obscured sources.
Since the diffuse flux is the sum of the contributions from all PWBs, 
its energy spectrum reflects the distribution of $\gamma_p$ 
among the different PWBs (see Fig. 1), and could teach us about this
distribution. The spectrum is typically quite different form that 
of other sources, which would make it easier to identify if it is detected.
The diffuse flux from the $sh$ mechanism is
$\epsilon_\nu^2\Phi_\nu\approx 10^{-7}f_p\;{\rm GeV\; cm^{-2}\;s^{-1}\;sr^{-1}}$,
and extends over $\epsilon_\nu\sim 10^8-10^{15}\;$eV.
The exact value of the diffuse flux for our 3 neutrino production 
mechanisms ($p\gamma$, $pp$ and $sh$), is somewhat 
uncertain; the largest uncertainty is for $pp$. 

Our modeling of the neutrino emission from PWBs that host GRBs 
also applies to normal PWBs, like the Crab or Vela, when they are very
young and have not spun down considerably, $t\lesssim t_{\rm sd}$. 
The main differences are that $E_{\rm rot}\sim 10^{51}\;$erg, $\beta_b\sim 0.01$, 
$t_{\rm sd}\sim 100\;$yr and $R_b\propto t$. Since there are $\sim 100$ 
times more regular PWBs, the total energy budget is comparable. 
Photomeson interactions become unimportant, but 
$a_{pp}=2$ and $f_\pi^{pp}\sim 1$ at $t\lesssim t_{pp}\sim 20\;$yr 
so that $f_\pi^{pp}$ time averaged over $0<t\lesssim t_{\rm sd}$ is 
$\langle f_\pi^{pp}\rangle\sim 2(t_{pp}/t_{\rm sd})\sim 0.4$,
and neutrino production via p-p collisions is rather efficient.
The diffuse flux from normal PWBs is therefore expected to be comparable, 
to that from PWBs that host GRBs. 
The neutrino emission from normal PWBs was recently calculated~\cite{BB}
assuming larger proton energies, resulting in a diffuse neutrino spectrum 
extending to higher energies but with similar 
$\epsilon_\nu^{2}(dN/d\epsilon_\nu)$ peak flux levels. 
Since normal PWBs are much more common, galactic PWBs like the 
Crab and Vela might be detected by planned ${\rm km^2}$ detectors even 
though $t>t_{\rm sd}$ for these sources~\cite{normal_PWBs}.

In order for GRBs to have typical afterglow emission, we need 
$t_{\rm sd}\gtrsim 20\;$yr, $\gamma_p\lesssim 10^5$ 
(for spherical PWBs)~\cite{GG},
implying $f_\pi^{p\gamma}<f_\pi^{pp}$, 
$\langle f_\pi^{pp}\rangle\lesssim 0.25$ and 
$N_\mu\sim 0.01\;{\rm km^{-2}}$
events per PWB. The expected detection rate of $\nu_\mu\bar\nu_\mu$ from all PWBs 
hosting GRBs with typical afterglows, that are detectable in $\gamma$-rays
(i.e. pointed towards us), is $\sim 10\;{\rm km^{-2}\; yr^{-1}}$. However, 
most of these $\nu$'s will arrive $\sim 20(1+z)\,$yr before the GRB, 
at $0<t\lesssim 4(1+z)\,$yr. Around the time of the GRB $f_\pi^{pp}\sim 10^{-3}$ 
and the rate of events is $N_\mu\sim 10^{-4}\;{\rm km^{-2}\; yr^{-1}}$.
Since the $\nu$'s from the PWB are emitted isotropically, the 
diffuse neutrino flux from these sources,  
$\epsilon_\nu^2\Phi_\nu\sim 2\times 10^{-7}\;{\rm GeV\; cm^{-2}\;s^{-1}\;sr^{-1}}$,
is dominated by GRBs that point away from us.
This diffuse flux is somewhat lower than that expected for PWBs with 
$t_{\rm sd}\lesssim 1\;$yr and normal PWBs,  
and is below the atmospheric neutrino background
(since $\gamma_p\lesssim 10^5$). 

The expected number of events per PWB with $t_{\rm sd}\lesssim 1\;$yr at $z\sim 1$ 
is $\sim 0.1\;{\rm km^{-2}}$, with neutrino energies $\epsilon_\nu\sim 1-10^3\;$TeV. 
These $\nu$'s are emitted over a time $t_{\rm sd}\sim 0.01-1\;$yr before the GRB,
and the emission decays over a similar time after the GRB. This emission is 
therefore easily distinguishable from $\nu$'s emitted either simultaneously 
with the GRB~\cite{GRB_nu} or $\lesssim 100\;$s before the GRB~\cite{coll_nu}, 
as predicted for the collapsar model of GRBs. 
The number of neutrino events from the PWB is at least an order of magnitude 
larger than from the prompt GRB, and is much larger than from the 
afterglow~\cite{WB00}. Therefore, this GRB precursor neutrino signal is one of 
the best candidates for an early detection with the planned ${\rm km^2}$ 
neutrino telescopes~\cite{Halzen}. The detection of a neutrino signal from a 
PWB can serve to distinguish between the different progenitor models for GRBs.

\paragraph*{Acknowledgments.}

We thank P. Goldreich, A. K\"onigl, E. Waxman, A. Loeb, P. M\'esz\'aros
and C. Pe\~na-Garay for useful discussions.
JG is supported by the Institute for Advanced Study, 
funds for natural sciences.
DG thanks the IAS in Princeton, for the hospitality and the 
nice working atmosphere during the course of this work.

\end{document}